\documentstyle[times,prl,aps,epsf,balanced]{revtex}
\begin{document}

\title{Disordering effects of color in nonequilibrium phase
transitions\\ induced by multiplicative noise}

\author{S. Mangioni$^1$, R. Deza$^1$, H.S. Wio$^2$ and R.Toral$^3$}

\address{(1) Departamento de F\'{\i}sica, Facultad de Ciencias
Exactas y Naturales \\ Universidad Nacional de Mar del Plata, De\'an
Funes 3350, 7600 Mar del Plata, Argentina.\\
(2) Centro At\'omico Bariloche (CNEA) and Instituto Balseiro (U.N. de
Cuyo), 8400 S.C. de Bariloche, Argentina.\\ (3) Departament de
F\'{\i}sica, Universitat de les Illes Balears and Instituto
Mediterr\'aneo de \\Estudios Avanzados, IMEDEA (CSIC-UIB), E-07071
Palma de Mallorca, Spain}

\maketitle

\begin{abstract}
The model introduced by Van den Broeck, Parrondo and Toral [Phys.\
Rev.\ Lett.\ {\bf 73}, 3395 (1994)]---leading to a second-order-like
{\em noise-induced nonequilibrium phase transition\/} which shows {\em
reentrance\/} as a function of the (multiplicative) noise intensity
$\sigma$---is investigated beyond the white-noise assumption.  Through
a Markovian approximation and within a mean-field treatment it is
found that---in striking contrast with the usual behavior for
equilibrium phase transitions---for noise self-correlation time
$\tau>0$, the {\em stable\/} phase for (diffusive) spatial coupling
$D\to\infty$ is always the {\em disordered\/} one.  Another surprising
result is that a large noise ``memory'' also tends to {\em destroy\/}
order.  These results are supported by numerical simulations.
\end{abstract}


\begin{twocolumns}
%\begin{twocolumn}
Too often do we resort, in studying nonequilibrium systems, to the
paradigmatic body we have inherited from equilibrium thermodynamics.
Though most times this way of reasoning is of valuable help for us to
interpret the results, we should be more aware of the fact that
sometimes it can be seriously misleading.  An archetypical example is
the intuitive image we have developed of a close relationship between
{\em noise\/} and {\em disorder\/}, and between {\em spatial
coupling\/} (and also between {\em time-correlation\/}) and {\em
order\/}.  Regarding the first, whereas it is true that studies on
e.g.\ Ginzburg-Landau models subject to {\em additive\/} noise seem to
reinforce this ``rule'',\cite{to90,ga92,ga94} in the last decade we
have also witnessed examples of exactly the opposite trend---namely,
dynamical systems in which a {\em multiplicative\/} noise couples to
the system's nonlinearities in such a way that it generates a
transition towards an {\em ordered\/} state.  In fact, it is by now a
well-known fact that the noise can induce a unimodal-bimodal
transition in some $0$-dimensional models;\cite{HL} nevertheless, this
result can still be argued to be somewhat restricted since in this
case there cannot be breakdown of ergodicity, which is required for a
true (nonequilibrium, noise-induced) {\em phase\/} transition.

Recently, a model was introduced whereby an {\em extended\/} system
subject to a Gaussian multiplicative noise---white both in space and
time---can undergo a noise-induced symmetry-breaking transition
towards an ordered state: this became the first example of a {\em
purely\/} noise-induced, nonequilibrium, ordering {\em phase\/}
transition.\cite{VPT} This result was obtained within a mean-field
approximation and confirmed afterwards through extensive simulations
in $d=2$.\cite{VPTK} In this case, and at variance with the case of
order-disorder transitions at equilibrium (induced as we know by the
spatial coupling constant $D$ and the bistability of the local
potential) it is the short-time instability induced by the
multiplicative noise intensity $\sigma$---reinforced by the spatial
coupling $D$---which induces the transition.  Neither the $d=0$ system
($D=0$) nor the deterministic one ($\sigma=0$) show any transition;
moreover---and strikingly enough---those systems exhibiting
noise-induced transitions in $d=0$ are automatically ruled out as
candidates for this phenomenon.  Such a noise-induced phase
transition---besides being of a second-order type as a function of the
noise intensity---has the noteworthy feature of being {\em reentrant}:
the ordered state can be found only inside a window determined by two
values of $\sigma$.  A similar reentrant effect has been observed in
the Ginzburg-Landau model with multiplicative and additive
noises.\cite{GOPSB}

One can question whether it is realistic enough to consider a genuine
multiplicative noise as white.  It appears more likely that the kind
of fluctuations leading to multiplicative noise---coming in general
from the coupling with an external source---will exhibit some degree
of spatiotemporal correlation.\cite{HL,S,LHK,SS} Moreover, one should
expect new nontrivial effects as a consequence of the color of the
multiplicative noise: after all---whereas it was shown in
Refs.\cite{ga92,ga94} that an {\em ordering\/} nonequilibrium phase
transition can be induced in a Ginzburg-Landau model by varying the
correlation time of the {\em additive\/} noise---a {\em reentrant\/}
behavior has been found recently in a ($d=0$) colored-noise-induced
transition.\cite{CSW}

It is our aim in this work to investigate the effects of the
self-correlation time $\tau$ of the multiplicative noise on the model
of Refs.\cite{VPT,VPTK}.  To that end we use an Ornstein-Uhlenbeck
(OU) noise and apply---in the framework of a mean-field treatment---a
``unified colored-noise approximation" (UCNA),\cite{JH} together with
an interpolation scheme that extends its range of validity in
$\tau$.\cite{CWA} Our main finding is that---at variance with the
usual behavior in {\em equilibrium\/} statistical mechanics---a large
coupling constant $D$ leads invariably for $\tau>0$ to a {\em
disordered\/} state.  Since (as discussed thoroughly in
Ref.\cite{VPTK}) the clue for the phase transition seems to reside in
an instability occurring in the short-time behavior, and the model
introduced in Ref.\cite{VPT} was precisely chosen as a representative
(perhaps the simplest one) of a host of systems exhibiting such an
instability---and hopely, a noise-induced phase transition with
similar characteristics---this unexpected result warns
experimentalists seeking for concrete realizations of this phenomenon
not to tune naively the spatial coupling intensity $D$ up to a very
large value (as one would do guided by ``experience'') but to look
instead for an optimal value of $D$ for which the order parameter
would take its maximum value.

Another striking result is that---consistently with the result in
Ref.\cite{CSW}---increasing the ``memory'', i.e.\ the self-correlation
time $\tau$ of the noise, does not favor (as one would naively expect)
the transition towards an ordered phase but all the way around.  This
is indicated by the following facts: (a) the threshold (critical)
value of $\sigma$ is a strongly increasing function of $\tau$, and (b)
the window in $D$ available to the ordered phase strongly shrinks as
$\tau$ increases.  There is nonetheless a hint that (like in
Ref.\cite{CSW}) a small amount of color could induce order slightly
beyond the upper critical value of $\sigma$ corresponding to $\tau=0$,
but since the region in which this phenomenon occurs is somewhat
narrow, and the comparison with simulations in $d=2$---together with
finite-size scaling---made in Refs.\cite{VPT,VPTK} sheds much doubt on
the precise location of this value, we prefer not to take this result
too seriously for the moment (although it certainly deserves further
analysis).

As in Ref.\cite{VPT} we shall resort to a lattice version of the
extended system, whose state at time $t$ will then be given by the
set of stochastic variables $\{x_i(t)\}$ ($i=1,\dots,L^d$) defined at
the sites ${\bf r}_i$ of a hypercubic $d$-dimensional lattice of side
$L$. The variables $\{x_i\}$ obey the following system of ordinary
stochastic differential equations (SDE):
\begin{equation}\label{1}
{\dot x}_i=f(x_i)+g(x_i)\eta_i+\frac{D}{2d} \sum_{j\in n(i)}(x_j-x_i)
\end{equation}
where $D$ is the lattice version of the diffusion coefficient, $n(i)$
stands for the set of $2d$ sites which form the immediate
neighborhood of site ${\bf r}_i$, and $\eta_i$ is the {\em colored
multiplicative\/} noise acting on site ${\bf r}_i$. This coupled set
of Langevin-like equations is the discrete version of the {\em
partial\/} SDE which in the continuum would determine the state of
the extended system, the last term being replaced---in the continuum
limit---by the Laplacian operator $\nabla^2 x$. The specific case
analyzed in Ref.\cite{VPT} (which the authors conjecture that could
be the simplest example exhibiting such a transition) is
\begin{equation}
f(x)=-x(1+x^2)^2\quad{\rm and}\quad g(x)=1+x^2 \end{equation}
As in Refs.\cite{ga92,CSW}, the noises $\{\eta_i\}$ are taken to be
OU ones, i.e.\ Gaussian-distributed stochastic variables with zero
mean and the following correlations: \begin{equation}
\langle\eta_i(t)\eta_j(t')\rangle=\delta_{ij}
\frac{\sigma^2}{2\tau}\exp\left(-\frac{|t-t'|}{\tau}\right)
\end{equation}
In the limit $\tau\to 0$ the OU noise $\eta_i(t)$ tends to the
white noise $\xi^W_i(t)$ with correlations
$\langle\xi^W_i(t)\xi^W_j(t')\rangle=\sigma^2\delta_{ij}\delta(t-t')$,
which is the case studied in Ref.\cite{VPT}.

The non-Markovian character of the process $\{x_i\}$ (due to the
colored noise $\{\eta_i\}$) makes it difficult to study.  However,
there are some approximate Markovian techniques that---whereas
capturing some of the essential features of the complete non-Markovian
process---strongly simplify the treatment of the equations, allowing
to exploit well-known Markovian techniques.\cite{SS} Amongst those
approximations, the UCNA and related interpolation schemes are very
useful since they can reproduce the limits of small and large
correlation time $\tau$.\cite{JH,CWA} As discussed in Ref.\cite{JH}
for a single SDE, the conditions assumed in the UCNA indicate that its
validity should decrease with increasing noise intensity.  On the
other hand, regarding the $\tau$-dependence, the UCNA becomes {\em
exact\/} for $\tau\to 0$ and for $\tau\to\infty$.  Although the
interpolation procedure in Ref.\cite{CWA} extends the validity range
of this effective Markovian approximation, it is still not clear how
far it does so.

We now sketch the main lines of our calculation (a more detailed
account is given in Ref.\cite{MDWT}):\\
i) For our particular problem, the UCNA proceeds by taking the time
derivative of Eqs.(\ref{1}) and---after substitution of the Langevin
equations satisfied by the OU noise ($\tau\dot\eta_i=-\eta_i
+\sigma\xi_i$, where $\{\xi_i\}$ are standard white-noise
variables)---setting to zero not only $\ddot x_i$ (a usual adiabatic
elimination) but also $(\dot x_i)^2$, in order to recover a proper
Fokker-Planck equation (FPE) description.\cite{WCSPR}\\
ii) In order to reduce the complexity of the resulting system of
Markovian SDE, we make the {\em approximation\/} of replacing (under
the hypothesis that the system is isotropic) {\em in each equation\/}
of that set the $2d$ variables $x_j$ by a single one $y_i$.\\
iii) Within the Stratonovich prescription we are left with the FPE for
a bivariate steady-state probability distribution function (pdf)
$P^{st}(x_i,y_i)$.  To the drift and diffusion coefficients of this
FPE we apply an approximation in the spirit of the Curie-Weiss
mean-field type of approach used in Ref.\cite{VPT}, so deriving an
effective stationary joint pdf $P^{st}(x,y)$ (we have dropped the
subindex $i$ for brevity), from which we derive a one-site pdf
$P^{st}(x;\langle x\rangle)$ by assuming
$P^{st}(x,y)=P^{st}(x)\,\delta(y-\langle x\rangle)$.\\
iv) The value of $\langle x\rangle$ follows then from a
self-consistency relation similar to that of Ref.\cite{VPT}:
\begin{equation}\label{cons}
	\langle x\rangle=\int dx\,x P^{st}(x;\langle x\rangle)
\end{equation}
This equation has always the trivial solution $\langle x\rangle=0$
corresponding to a disordered phase.  When other stable, nontrivial,
$\langle x\rangle\ne 0$ solutions appear, the system develops order
through a genuine phase transition and $m\equiv|\langle x\rangle|$ can
be considered as the order parameter (due to the symmetry of the
problem, both $\pm \langle x\rangle$ are solutions of the previous
equation).  In the white-noise limit $\tau=0$ this is known to be the
case for sufficiently large values of the coupling $D$ and for a
window of values of the noise intensity
$\sigma\in[\sigma_1,\sigma_2]$.

We now discuss how the presence of ordered states is altered by
nonzero values of $\tau$ in the mean-field study.  Figure \ref{fig1}
shows, in the parameter subspace $\sigma-D$, the boundaries separating
the ordered and disordered phases for different values of $\tau$.  The
noteworthy aspects of this graph are the following:\\
i) For fixed $\sigma>1$ and $\tau>0$ the ordered states can exist {\em
only within a window\/} of values for $D$.  In other words, the
noise-induced nonequilibrium phase transition exhibits reentrance not
only with respect to $\sigma$ (as in the $\tau=0$ case) but also with
respect to $D$.\\
ii) For fixed $D$ and $\sigma$ inside the $\tau=0$ phase boundary---as
indicated, for example, by point (a) in Fig.\ref{fig1}---there always
exists a value of the correlation time $\tau$ beyond which the system
becomes disordered.  Furthermore, there seems to exist a value of
$\tau>0.123$ beyond which order is impossible, whatever the values of
$\sigma$ and $D$.\\
iii) For fixed (and large enough) values of $D$, and for values of
$\sigma$ that would correspond to the disordered phase for $\tau=0$, a
small increase in $\tau$ induces a transition towards an ordered
phase---as indicated by the point marked (b) in Fig.\ref{fig1}.
However, a further increase in $\tau$ can again lead to disorder.  In
other words, the transition can also be reentrant with respect to
$\tau$.  Regarding the reentrant nature of the transition with respect
to $D$, in Fig.\ref{fig1} it can be seen that---as $\tau$ increases
from zero---the maximum value of $D$ compatible with the ordered phase
reaches, for $\sigma$ large enough, a ``plateau'' which is a
decreasing function of $\tau$.  At the same time, the minimum value of
$D$ (that at $\tau=0$ goes like $D\propto\sigma_2^2$) tends also to
become constant as a function of $\sigma$ as $\tau$ increases, so
shrinking the window available for the ordered phase until it
virtually disappears.
\begin{figure}[h]
\begin{center}
\def\epsfsize#1#2{0.46\textwidth}
\leavevmode
\epsffile{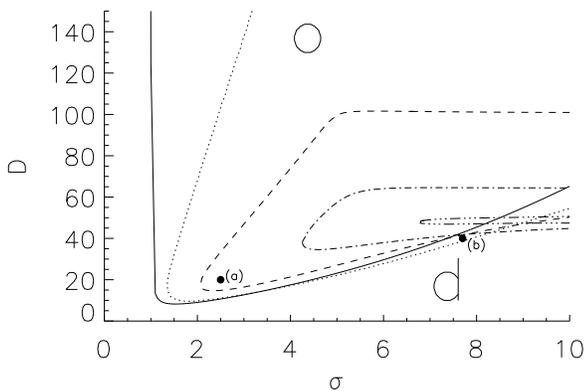}
\end{center}
\caption{\label{fig1} Projection of the mean-field phase diagram onto
the $\sigma-D$ plane, for $\tau=0$ (continous line), $\tau=0.015$
(dotted line), $\tau=0.05$ (dashed line), $\tau=0.1$ (dotted-dashed
line), and $\tau=0.123$ (triple dotted-dashed line).  For each curve,
the ordered zone is the area inside the curve (for $\tau=0$ we have
marked the ordered and disordered regions with ``o'' and ``d'',
respectively).  Points (a) and (b) correspond to the transitions
referred to in the text.}
\end{figure}
Regarding the character of the transition at $\tau=0$, we have
checked that as $\tau$ decreases the lower-left elbow climbs up the
$\sigma_1$ branch corresponding to $\tau=0$, but the slope of the
corresponding $\tau\neq 0$ branch is always positive.

The previous features can also be inferred from the behavior of the
order parameter $m$. Plotting $m$ vs.\ $\sigma$ for a fixed value of
$D$ and different values of $\tau$, we would see a general trend of
the ordered zone to shrink and disappear with increasing $\tau$.
Whereas the lower critical value $\sigma_1$ increases monotonically
with $\tau$, the upper value $\sigma_2$
first increases a little and then becomes a monotonically decreasing
function of $\tau$. This is, of course, consistent with what occurs
around point (b) in Fig.\ref{fig1}.
\begin{figure}[h]
\begin{center}
\def\epsfsize#1#2{0.46\textwidth}
\leavevmode
\epsffile{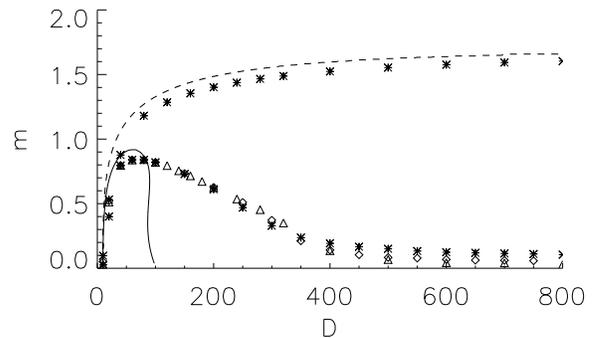}
\end{center}
\caption{\label{fig2} Mean-field prediction for the order parameter
$m$ as a function of the spatial coupling $D$, for noise intensity
$\sigma=2$ and self-correlation times $\tau=0$ (dashed line) and
$\tau=0.01$ (continuous line).  Notice that whereas for $\tau=0$ the
curve tends to the asymptotic value $(\sigma^2-1)^{1/2}=1.73$, for
$\tau=0.01$ the order parameter seems to fall off rather abruptly to
zero for a large value $D_{MF}$ of the coupling.  Simulation results
for different system sizes: $L=16$ (asterisks), $L=32$ (rhombs) and
$L=64$ (triangles) are also included, showing that $m$ indeed decays
to zero, although much slowly and for a much larger ``critical'' value
$D_c$.}
\end{figure}
Since the previous results have been obtained in the mean-field and
UCNA approximations, and their range of validity is somewhat unclear,
we have also performed numerical simulations in order to have an
independent check of the predictions. As a representative example
(corresponding to the phenomenon (i) above---namely, the destruction
of the ordered phase by an increasing coupling constant $D$) we plot
in Fig.\ref{fig2} $m$ vs.\ $D$ as predicted by our mean-field theory,
and results coming from a numerical integration of the SDE, for
$\sigma$ fixed and two values of $\tau$. Although for $\tau\neq 0$
the numerical results do not follow the mean-field theory, it is
obvious that there is an optimal value of the coupling $D$ for which
the order parameter takes a maximum value, and that order {\em
disappears\/} for $D$ large enough. From Fig.\ref{fig2} one cannot
decide whether the maxima will accompany the $\tau=0$ curve as
$\tau\to 0$. It could well be that the phase transition at $\tau=0$
for $\sigma$ fixed and $D$ large enough be even a {\em first-order\/}
one. This certainly calls for further investigation.

We stress again the fact that these effects of a colored
multiplicative noise on an extended dynamical system (unable to
undergo any phase transition in the absence of noise) are {\em
qualitatively\/} different to the ones observed in (nonequilibrium)
phase transitions driven by a colored additive noise on a prototypic
model for equilibrium phase transitions.\cite{ga92,ga94} Whereas in
the last case the role of the correlation time is to {\em
stabilize\/} the ordered phase and/or induce order in systems that
are disordered for $\tau=0$, the main effect of color in our case is
to {\em destroy\/} order. Also, we should not be turned back by the
quantitative disagreement between the mean-field theory and the
numerical simulations: it is known that in equilibrium phase
transitions, mean-field theory overestimates the ordered region
and---for example---in the previous study of the same model with
white noise,\cite{VPT,VPTK} the mean-field prediction for the upper
critical value $\sigma_2$ for the reentrant transition was thrice the
one found numerically. Although the numerical results are affected by
finite-size effects---as one would expect in a second-order phase
transition---one can see unambiguously in Fig.\ref{fig2} the decrease
of the order parameter with increasing coupling $D$, for $\tau$ as
small as $0.01$.

In order to understand this sudden change in behavior as soon as a
tiny self-correlation is present, we have studied the time evolution
equation for $\langle x\rangle$ (which is small when the parameters
are around the phase boundary) within the mean-field approximation,
as $\frac{D}{\sigma^2}\to\infty$. In Ref.\cite{VPTK} this simple {\em
linear\/} (i.e., up to first-order in $\langle x\rangle$) criterion
of stabilisation of the disordered phase was introduced as a way of
determining the region of appearance of ordered phases. For $\tau\ll
1$ and $\frac{D}{\sigma^2}\to\infty$ (we indeed assume $\tau D$ and
$\sigma^2$ to be $O(1)$), it reads \begin{equation}\label{2}
\langle\dot x\rangle=-\alpha\langle x\rangle,\quad{\rm with}\quad
\alpha=\frac{1+\tau D-\sigma^2}{1+\tau D}. \end{equation}
When $1+ \tau D>\sigma^2$ it is $\alpha>0$, and hence the disordered
phase ($\langle x\rangle=0$) is stable. On the other hand, if $1+\tau
D<\sigma^2$ it is $\alpha<0$, and it is the ordered phase ($\langle
x\rangle\neq 0$) which becomes stable. In summary, whereas the noise
intensity $\sigma$ has a stabilizing effect on the ordered phase, as
soon as $\tau\neq 0$ the spatial coupling $D$ tends to {\em
destabilize\/} it. For $\tau=0$ the last effect is not present, being
then the condition for ordering that $\sigma>1$ (this is the effect
that was reported in Refs.\cite{VPT,VPTK}). Considering that the
effect of even a tiny correlation is enhanced by $D$, we can
understand the abrupt change shown in Figs.\ref{fig1} and \ref{fig2}
as soon as $\tau\neq 0$.

This work has focused on the effects of a self-correlation in the
multiplicative noise on the reentrant noise-induced phase transition
reported in Ref.\cite{VPT}. It appears that for $\tau\neq 0$, a
strong enough spatial coupling is capable of destroying the order
established as a consequence of the multiplicative character of the
noise.
The foregoing result can be understood by recalling the fact that the
ordered phase arises as a consequence of the collaboration between
the multiplicative character of the noise and the presence of spatial
coupling. When no self-correlation is present, the disordering effect
of $D$ cannot be felt. This explains the results in Ref.\cite{VPTK},
which have been rightly interpreted in terms of a ``freezing'' of the
short-time behavior by a strong enough spatial coupling. As $\tau$
increases, the minimum value of $D$ required to destabilize the
ordered phase becomes lower and lower. In this way, the region in
parameter space available to the ordered phase shrinks further and
further until it vanishes.

The main lesson one can draw from the present results is that the
conceptual inheritance from equilibrium thermodynamics (though often
useful) is not always applicable. By following the
equilibrium-thermodynamic lore, one would tend to think that as
$D\to\infty$ an ordered situation is favored. This is certainly true
for the Curie-Weiss-type models, since in that case the deterministic
potential is itself bistable and an increase of spatial coupling has
the effect of rising the potential barrier between the stable states.
In the case we are dealing with, the deterministic potential is
monostable and it is the combined effects of the multiplicative noise
and the spatial coupling that induce the transition.

R. Toral acknowledges financial support from DGICyT, projects number
PB94-1167 and PB94-1172 and H. S. Wio from Fundaci\'on Antorchas,
project A-13396/1 and CONICET PIP-4593/96.

\end{twocolumns}
%\end{twocolumn}

\begin{references}
\bibitem{to90} R. Toral and A. Chakrabarti, Phys.\ Rev.\ B {\bf 42},
2445 (1990) and references therein.

\bibitem{ga92} J. Garc\'{\i}a-Ojalvo, J. M. Sancho and L.
Ram\'{\i}rez-Piscina, Phys.\ Lett.\ A {\bf 168}, 35 (1992).

\bibitem{ga94} J. Garc\'{\i}a-Ojalvo and J. M. Sancho, Phys.\ Rev.\ E
{\bf 49}, 2769 (1994).

\bibitem{HL} W. Horsthemke and R. Lefever, {\it Noise--Induced
Transitions: Theory and Applications in Physics, Chemistry and
Biology} (Springer, 1984).

\bibitem{VPT} C. Van den Broeck, J. M. R. Parrondo and R. Toral,
Phys.\ Rev.\ Lett.\ {\bf 73}, 3395 (1994).

\bibitem{VPTK} C. Van den Broeck, J. M. R. Parrondo, R. Toral and R.
Kawai, Phys.\ Rev.\ E {\bf 55}, 4084 (1997).

\bibitem{GOPSB} J. Garc\'{\i}a-Ojalvo, J. M. R. Parrondo, J. M.
Sancho and C. Van den Broeck, in Phys.\ Rev.\ E {\bf 54}, 6918
(1996).

\bibitem{S} A. V. Soldatov, Mod.\ Phys.\ Lett.\ B {\bf 7}, 1253
(1993).

\bibitem{LHK} I. L'Heureux and R. Kapral, J. Chem.\ Phys.\ {\bf 90},
2453 (1988).

\bibitem{SS} J. M. Sancho and M. San Miguel, in {\it Noise in
Nonlinear Dynamical Systems}, F. Moss and P. V. E. McClintock, eds.\
(Cambridge U. Press, 1989), p.72.

\bibitem{CSW} F. Castro, A. D. S\'anchez and H. S. Wio, Phys.\ Rev.\
Lett.\ {\bf 75}, 1691 (1995).

\bibitem{JH} P. H\"anggi and P. Jung, ``Colored Noise in Dynamical
Systems'', in {\it Advances in Chemical Physics} vol.LXXXIX, I. 
Prigogine and S. A. Rice, eds.\ (J. Wiley, 1995), p.239.

\bibitem{CWA} F. Castro, H. S. Wio and G. Abramson, Phys.\ Rev.\ E
{\bf 52}, 159 (1995).

\bibitem{MDWT} S. Mangioni, R. Deza, H. S. Wio and R. Toral, to be
submitted to Phys.\ Rev.\ E.

\bibitem{WCSPR} H. S. Wio, P. Colet, M. San Miguel, L. Pesquera and
M. Rodr\'{\i}guez, Phys.\ Rev.\ A {\bf 40}, 7312 (1989).

\end{references}
\end{document}